\documentclass[%
 reprint,
nofootinbib,
 amsmath,amssymb,
 aps,
]{revtex4-1}

\usepackage{color}
\usepackage{textcomp}
\usepackage{hyperref}
\usepackage{graphicx}
\usepackage{dcolumn}
\usepackage{bm}

\usepackage{tikz}

\newcommand{\customyinyang}[1][1]{%
	\begin{tikzpicture}[scale=#1*0.07]
	\draw[line width = #1*0.05mm,transform canvas={yshift=0.02cm}] (0,0) circle (1cm);
	\path[fill=black,transform canvas={yshift=0.02cm}] (90:1cm) arc (90:-90:0.5cm)
	(0,0)    arc (90:270:0.5cm)
	(0,-1cm) arc (-90:-270:1cm);
	
	\end{tikzpicture}}

\begin{document}

\preprint{APS/123-QED}

\title{Modularity maximization as a flexible and generic \\ framework for brain network exploratory analysis}

\author{Farnaz Zamani Esfahlani$^1$\textdagger}
\author{Youngheun Jo$^{1}$\textdagger}
\author{Maria Grazia Puxeddu$^{1-3}$\textdagger}
\author{Haily Merritt$^4$\textdagger}
\author{Jacob C. Tanner$^4$\textdagger}
%\author{Brock Glaser$^{1\customyinyang[1]}$}
\author{Sarah Greenwell$^{1\customyinyang[1]}$}
%\author{Luis Nieves$^{1\customyinyang[1]}$}
\author{Riya Patel$^{1\customyinyang[1]}$}
%\author{Jonah Slack$^{1\customyinyang[1]}$}
\author{Joshua Faskowitz$^{1,5}$}
\author{Richard F. Betzel$^{1,4-6}$}
\email{rbetzel @ indiana.edu}

\affiliation{
	$^1$Department of Psychological and Brain Sciences, $^4$Cognitive Science Program, $^5$Program in Neuroscience, $^6$Network Science Institute, Indiana University, Bloomington, IN 47405
}

\affiliation{
	$^2$Department of Computer, Control and Management Engineering ``Antonio Ruberti'', Sapienza University of Rome, 00185 Rome, Italy,
	$^3$IRCCS Fondazione Santa Lucia, 00185 Rome, Italy, Italy
}

\affiliation{\textdagger = contributed equally}

\date{\today}
\begin{abstract}
The modular structure of brain networks supports specialized information processing, complex dynamics, and cost-efficient spatial embedding. Inter-individual variation in modular structure has been linked to differences in performance, disease, and development. There exist many data-driven methods for detecting and comparing modular structure, the most popular of which is modularity maximization. Although modularity maximization is a general framework that can be modified and reparamaterized to address domain-specific research questions, its application to neuroscientific datasets has, thus far, been narrow. Here, we highlight several strategies in which the ``out-of-the-box'' version of modularity maximization can be extended to address questions specific to neuroscience. First, we present approaches for detecting ``space-independent'' modules and for applying modularity maximization to signed matrices. Next, we show that the modularity maximization frame is well-suited for detecting task- and condition-specific modules. Finally, we highlight the role of multi-layer models in detecting and tracking modules across time, tasks, subjects, and modalities. In summary, modularity maximization is a flexible and general framework that can be adapted to detect modular structure resulting from a wide range of hypotheses. This article highlights opens multiple frontiers for future research and applications.

\end{abstract}

\maketitle
\section*{Introduction}

The human brain is fundamentally a network and can be described at multiple spatiotemporal scales in terms of nodes -- neural elements, e.g. cells, populations,  areas -- and edges, e.g. synapses, projections, fiber tracts, and correlated activity \cite{park2013structural, sporns2005human}. Over the past two decades, network analyses have uncovered key attributes of brain networks that are believed to support distinct modes of brain function. These include small-world architecture to support segregated and specialized information processing and efficient integration of information over long distances \cite{bassett2006small,sporns2004small} , influential and central hubs that form integrative rich clubs \cite{hagmann2008mapping,van2011rich}, and cost-efficient spatial embeddings \cite{bassett2010efficient}.

Additionally, brain networks can be decomposed into subgraphs -- referred to as modules or communities in the language of network science \cite{sporns2016modular,meunier2009hierarchical}. Modules are typically defined as cohesive subgraphs, where neural elements within the same modules tend to form strong connections to one another while those in different modules are weakly connected or not connected at all. This type of network organization helps support specialized processing, has been used to delineate the brain's functional systems \cite{power2011functional,gordon2016generation}, and is thought to support human cognition \cite{medaglia2015cognitive}. Studying the brain's modular structure confers obvious advantages. It helps with dimension reduction \cite{betzel2017modular}, facilitates discovery of underlying structure, and allows for classification of nodes' roles \cite{guimera2005functional}. 

How does one actually go about studying a network's community structure? On the one hand, network neuroscientists can borrow from one of any number of previously discovered community partitions to assign nodes (parcels) to communities or cognitive systems. For instance, \cite{power2011functional} describes a division of the brain into thirteen systems based on their network analysis. Other studies have reported similar divisions \cite{gordon2016generation, schaefer2018local}, allowing users to obtain a validated set of assignments for brains in their study.

However, there are many cases where we might expect previously defined communities to be unsuitable, for instance, in case-control studies, where we anticipate ahead of time that a clinical population might exhibit communities dissimilar from those of a neurotypical individual \cite{alexander2010disrupted}. More generally, if we want to capture individual differences in community structure, then imposing a shared partition onto a cohort may not be especially useful.

In these cases, we can use data-driven methods to algorithmically discover a network's community structure \cite{fortunato2010community,fortunato2016community,newman2012communities}. In most cases, these algorithms aim to partition a network's nodes into clusters that maximize some objective function. For instance, Infomap seeks a maximally compressible description of a random walker moving over a network's edges \cite{rosvall2008maps, sanchez2021detecting}. Stochastic blockmodels maximize the likelihood that a probabilistic model generated the observed network \cite{peixoto2014hierarchical}. Among this class of data-driven algorithms, however, modularity maximization is among the best known and most widely applied within network neuroscience \cite{newman2004finding}. The aim of modularity maximization, in general, is to partition nodes into communities so that the observed density of connections within communities is maximally greater than what we would expect had the network been generated under some null connectivity model. This intuition is captured by the modularity objectivity function, $Q$.

Although this definition of communities is general and can accommodate many distinct definitions of the ``null connectivity model'', its application in the context of network neuroscience has been narrow. In most applied studies, the null connectivity model generates random networks with a fixed degree and strength distribution. This ``enshrined'' model, which was promoted in the original studies of modularity, helps correct for communities that are driven by a network's degree distribution, i.e. groups of nodes that due to the fact that they make many connections are therefore more likely to be connected to one another. This model, however, has a number of issues -- e.g. it allows for self-connections, which are generally disallowed in the construction of empirical brain networks, and corresponds to a network with prohibitively expensive wiring cost.

The aim of this prospective article is not to provide a definitive roadmap for performing community detection or even how to use modularity maximization in practical contexts. For this, there are many useful reviews \cite{fortunato2016community,betzel2020community}. Rather, the aim of this article is to highlight modifications to a standard community detection algorithm that open up opportunities to address research questions that are central to network neuroscience and would be difficult to address otherwise.

In this study, we will examine several ways in which the modularity function could be extended to be more biologically plausible and to accommodate specific research questions and hypotheses. This work is divided into three sections. In the first section, we cover alternative null connectivity models and discuss their relative strengths and weaknesses, commenting on their suitability for network neuroscience research. In the second section, we show how modularity maximization can be used to discover clusters of brain regions that are dissimilar between conditions or correlated with a continuous variable. In the final section, we will introduce the multi-layer formulation for modularity maximization and highlight strategies for extending this framework to study multi-subject, multi-modal, and time-varying network datasets. Alongside this paper, we provide a documented toolbox so that researchers can apply these techniques to their own data.
 
\section*{Modularity maximization}

\subsection*{Network construction}

The human brain can be divided into functionally, cytoarchitectonically, and connectionally distinct regions or areas \cite{brodmann1909vergleichende, desikan2006automated, schaefer2018local}. These regions and their pairwise connections can be modeled as the nodes and edges of a graph or network and represented as its connectivity matrix, $A \in \mathbb{R}^{N \times N}$ whose element $A_{ij}$ encodes the existence and weight of the connection between nodes $i$ and $j$ \cite{bullmore2009complex, rubinov2010complex}. This representational framework can be used to model the network architecture of both functional and structural brain networks, although network inference and interpretation differs between modalities.

\subsubsection*{Structural connectivity}

Anatomical or structural connectivity (SC) refers to physical pathways that link neural elements to one another, including synaptic contacts \cite{white1986structure} and axonal projections \cite{oh2014mesoscale,markov2014weighted}. In the case of large-scale brain networks, these pathways represent white-matter fiber tracts which can be inferred noninvasively from diffusion-weighted MRI data using tractography algorithms \cite{sporns2005human,hagmann2008mapping}. SC changes over long timescales, e.g. with development \cite{baum2017modular,baum2020development} and aging \cite{betzel2014changes}, and is relatively invariant over shorter scales (the duration of a typical MRI scan). Although its precise density of connections varies across spatial scales and reconstruction procedure \cite{horvat2016spatial,ercsey2013predictive,sarwar2019mapping}, SC is generally sparse, meaning that many pairs of regions are not directly connected. This sparsity, along with the overall configuration of connections, imposes constraints on interregional communication patterns, effectively shaping the spatiotemporal evolution of brain activity \cite{honey2007network,adachi2012functional,avena2018communication}.

\subsubsection*{Functional connectivity}

Functional connectivity (FC), on the other hand, is a mathematical construct that measures the statistical relationship between activity recorded at two different points in the brain \cite{friston1994functional,horwitz2003elusive}. Although this definition includes virtually any bivariate (and possibly multivariate) statistical measure, FC is most commonly measured as the zero-lag correlation of fMRI BOLD activity between two regions (and usually extended to all pairs of regions). Whereas SC constrains interregional communication, FC reflects the patterns of correlated activity that emerge as a consequence of those constraints. Unlike SC, which changes across long timescales, FC evolves rapidly and fluctuates over short periods of time \cite{lurie2020questions,hutchison2013dynamic}.

\subsection*{Community structure of human brain networks}

Although SC and FC reflect distinct modes of interregional coupling, they both exhibit modular structure \cite{sporns2016modular}. Reported structural modules tend to be compact and contain spatially contiguous regions \cite{hagmann2008mapping}. The spatial contiguity of structural modules likely reflects evolutionary pressure to reduce the cost of wiring by forming short-range and hence less-costly connections \cite{stiso2018spatial}. Functional modules, because there is not explicit cost associated with the formation of long-distance correlations, are organized into brain-wide and spatially-distributed clusters \cite{yeo2011organization,power2011functional}. Even in the absence of explicit task instructions, functional modules recapitulate well-known patterns of task-evoked activity and previously-delineated functional systems \cite{smith2009correspondence}. These observations give rise to the hypothesis that modularity is a key component of functional specialization \cite{crossley2013cognitive}.

\subsection*{The modularity or ``Q'' heuristic}

\begin{figure*}[t]
	\centering
	\includegraphics[width=0.9\textwidth]{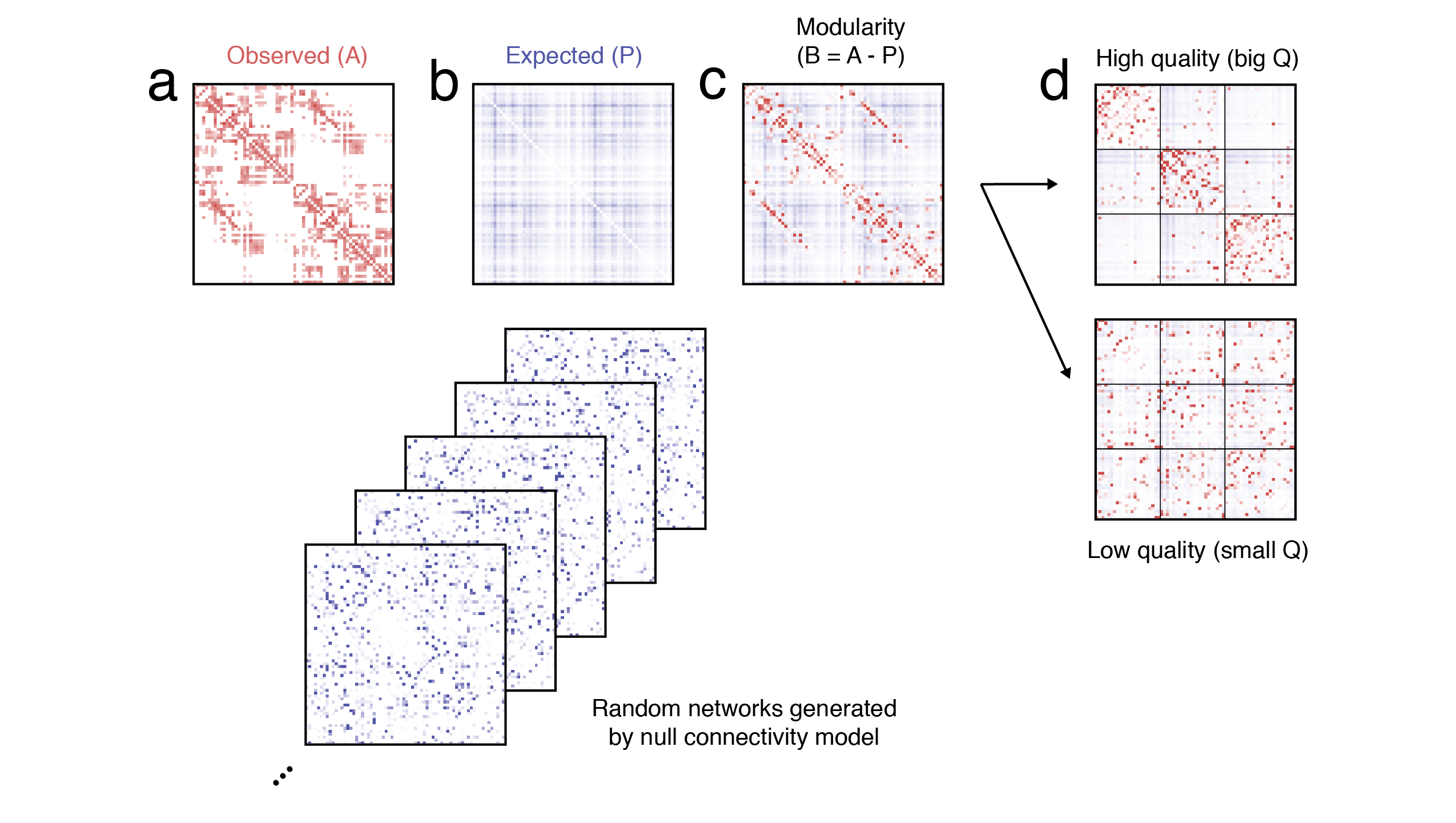}
	\caption{\textbf{Modularity maximization.} Modularity maximization works by comparing an observed network (\emph{a}) with what would be expected were one to simulate a null connectivity model an infinite number of times (\emph{b}). Specifically, the observed ($A$) and expected ($P$) weights of connections are compared to one another \emph{via} an element-wise subtraction, yielding a modularity matrix $B$ (\emph{c}). Modularity maximization aims to assign network nodes to communities so that the internal density of connections is maximally greater than chance. The modularity function, $Q$, is used to assess the quality of modular partitions. In panel \emph{d} we show examples of a high-quality partition (\emph{top}; within-community connections tend to be stronger than expected) and a low-quality partition (\emph{bottom}).} \label{figure+0}
\end{figure*}

In general, a network's modular structure is unknown ahead of time. Even in the case of SC and FC, whose modular structure has been characterized many times over, we may be interested in how the modular structure of individuals or groups of individuals deviates from composite, population-averaged modules. Assigning a network's nodes or edges to modules by hand or based on visual inspection would be tedious and unfeasible for large networks or even for small networks in large, multi-subject cohorts. Accordingly, the network elements are usually assigned to modules algorithmically through a process known as community detection \cite{fortunato2010community}.

There exists a multitude of algorithms and heuristics for detecting a network's modules \cite{lancichinetti2009community}. Some, like stochastic blockmodels, use statistical inference to group nodes into clusters with similar connectivity profiles. Most methods force nodes to belong to only one module (although others allow for overlap \cite{ahn2010link,evans2009line,faskowitz2020edge,palla2005uncovering}). Some methods detect modules based on how network properties shape the evolution of dynamical processes on the network \cite{rosvall2008maps,pons2006computing}. Others focus on the network's static topology alone.

In this last category is modularity maximization \cite{newman2004finding}. As noted in the introduction, modularity maximization operates on an eminently simple principle: compare what we actually see with what we might expect. In the context of networks, this means comparing an observed connectivity matrix, $A$ (Fig.~\ref{figure+0}a), with another matrix of identical dimensions, $P \in \mathbb{R}^{N \times N}$, whose element $P_{ij}$ encodes the weight of the connection between nodes $i$ and $j$ that we would expect under some null connectivity model (Fig.~\ref{figure+0}b). The simplest comparison of these matrices is an element-wise subtraction:

\begin{equation}
B = A - P.
\end{equation}

The resulting matrix, $B$, has a special name -- the \emph{modularity matrix} (Fig.~\ref{figure+0}c). The element $B_{ij}$ encodes whether the actual connection between $i$ and $j$ is stronger ($B_{ij} > 0$) or weaker ($B_{ij} < 0$) than we would expect. Modularity maximization uses the modularity matrix to evaluate the goodness or quality of a modular partition, i.e. a division of a network's nodes into non-overlapping modules. A partition's quality is rated by the modularity function, $Q$, which is calculated as:

\begin{equation}
Q = \sum_{ij} B_{ij} \delta(\sigma_i , \sigma_j).
\end{equation}

\noindent Here, $\sigma_i \in \{ 1 , \ldots , K  \}$ indicates to which of the $K$ modules that node $i$ is assigned and $\delta( \sigma_i , \sigma_j)$ is the Kronecker delta function, whose output is 1 when $\sigma_i = \sigma_j$ and 0 otherwise. Effectively, then, this double summation is over edges (node pairs) that fall within modules. The larger the value of $Q$, the higher quality the partition (Fig.~\ref{figure+0}d). That is, $Q$ tends to be large when communities are more internally dense than expected.

Rather than using $Q$ to simply evaluate the quality of a partition, $Q$ can be optimized outright, using a procedure known as modularity maximization. Like many clustering problems, discovering the partition that corresponds to the global maximum $Q$ is computationally intractable, and so many algorithms have been proposed to approximate its value. Among the most popular is the so-called ``Louvain algorithm'' -- a multi-stage, greedy algorithm that, in benchmark testing, is one of the fastest and most accurate \cite{blondel2008fast,jutla2011generalized} (although new versions challenge this \cite{traag2019louvain}). The Louvain algorithm is initialized with every node assigned to its own community. In random order, nodes are merged into larger communities if the merger yields an increase in $Q$. When no single-node moves can increase $Q$, nodes are aggregated into ``meta-nodes'' composed of all nodes with the same community label. The previous steps are then repeated until a complete cycle of single-node moves and aggregation yields no improvement.

\subsection*{Best practices}

Modularity maximization is a useful framework for discovering communities in a network. Using it effectively, however, requires that a user navigate several issues including the near-degeneracy of solutions as well as a potential resolution limit. Here, we offer some practical guidance for addressing these issues.

\subsubsection*{Near-degeneracy of the modularity landscape}

Obtaining the optimal partition of a network into communities is impossible in all but the most trivial instances. Stochastic methods like the Louvain algorithm generate estimates of this optimal partition, but it is understood that no single estimate should be treated as privileged. More generally, and this is especially true as the size of a network grows \cite{good2010performance}, there will be many dissimilar partitions of roughly the same quality. The reason for this is simple. Imagine we could obtain the globally optimal partition, i.e. the one corresponding to the largest possible $Q$. If we were to randomly select a node and move it to a new community, the partition would no longer be optimal, but it would likely still be of high quality. If we did this for every node, for every possible move, and then every pair of nodes, and every triplet, etc., we would obtain an ensemble of near-optimal solutions. In practice, methods like the Louvain algorithm sample from these types of partitions -- high quality, but likely sub-optimal. Importantly, there is heterogeneity among these solutions. So how do we solve this?

One possibility is to embrace this heterogeneity and focus on properties of this ensemble rather than those of any single partition \cite{peixoto2012entropy}. Another possibility is to condense information from the ensemble of partitions into a co-assignment matrix, whose elements indicate the fraction of partitions in which pairs of nodes were assigned to the same community \cite{kenett2020community}. This results in graded and fuzzy community assignments.

Oftentimes, however, it is useful to obtain a single representative partition to summarize the ensemble of near-optimal solutions. Again, there are multiple strategies for doing so. One possibility is to calculate the pairwise similarity of partitions to one another and select the one that is most similar, on average, to the others \cite{doron2012dynamic}. The most common strategy, however, is to estimate a consensus partition. In general, this involves iteratively reclustering the coassignment matrix \cite{lancichinetti2012consensus}, sometimes after applying a threshold to remove elements from node pairs that are infrequently coassigned to the same community. A potentially more useful and parameter-free method is to estimate, along with the empirical coassignment matrix, a matrix whose elements denote the expected co-assignment of nodes to communities, which can be easily obtained under a permutation model \cite{betzel2020community}. With the empirical and expected coassignment matrices, we can set up a modularity problem and use the modularity maximization heuristic to discover the consensus clusters.

\subsubsection*{Resolution limits and multi-scale extensions}

If we constructed a network of maximally connected cliques and linked cliques to one another by single edges, we would expect modularity maximization to discover the cliques as communities. In practice, however, for certain sized networks and null connectivity models, the presence of even a single link between communities may appear so ``unexpected'' that it is advantageous from the perspective of modularity maximization to group cliques together, defying the expectation that each clique correspond to a distinct module. This tendency to merge small, well-defined communities into larger clusters is known as the resolution limit \cite{fortunato2007resolution}. The implication is that, under certain conditions, modularity maximization may be blind to small communities.

To resolve this issue, many authors have incorporated a structural resolution parameter into the modularity equation. In its standard implementation, this parameter scales the relative contribution of the null connectivity model \cite{reichardt2006statistical, arenas2008analysis}. That is, $B = A - \gamma P$, where $\gamma$ is the resolution parameter. In effect, varying the value of $\gamma$ changes the scale (roughly, the size of communities) that modularity detects. When $\gamma$ is small, many elements of $A$ are greater than $\gamma P$ and, when grouped into large communities, have the net effect of increasing $Q$. On the other hand, when $\gamma$ is larger, a much smaller and exclusive set of connections in $A$ exceed $\gamma P$, yields much smaller communities. Thus, $\gamma$ acts as a way to zoom in and out, detecting finer and coarser structure, and helping to circumvent the resolution limit. In some instances, a ``sweep'' across $\gamma$ values, in which modular maximization is performed many times while systematically varying the resolution parameter, can be used to tune the results of modularity maximization to a specific number of communities or to obtain a multiresolution (or hierarchical) organization of the modular structure \cite{jeub2018multiresolution,harris2019hierarchical,akiki2019determining}.

\subsection*{Modifying modularity}

In practice, modularity maximization is used to address the question: ``here is a network, what are its modules?'' However, even this seemingly straightforward question requires that the user make a series of decisions \cite{betzel2020community}. Many of these decisions are fundamental and cannot be avoided even if modularity maximization were being used to address a different set of questions. These include how to deal with the degeneracy of the modularity landscape (the number of nearly-optimal partitions grows exponentially with the size of the network) \cite{good2010performance} and how to address issues of scale and the so-called resolution limit (under some conditions, modularity maximization can fail to detect small communities, even if they are unambiguously defined) \cite{fortunato2007resolution}. Other decisions are structural and concern how the modularity maximization problem is formulated. For instance, the user needs to define the expected weights of connections -- the variable $P$. This decision is non-trivial and will impact the character of the detected communities \cite{expert2011uncovering}.

The choice of null model is one decision that a user needs to make and is an example of how the modularity maximization problem can be modified. There are, of course, other ways to alter the way the problem is set up (e.g. by including a resolution parameter that scales the relative contribution of $P$ \cite{reichardt2006statistical,lambiotte2008laplacian,schaub2012markov}). In fact, any decision that changes the elements of the modularity matrix, $B$, has the capacity to impact the detected communities. In the following sections, we explore three strategies for modifying the modularity matrix and thereby changing the modularity maximization problem. We show that the modifications help address questions that are at the core of network neuroscience.

\section*{Single-layer modularity maximization}

\begin{figure*}[t]
	\centering
	\includegraphics[width=0.9\textwidth]{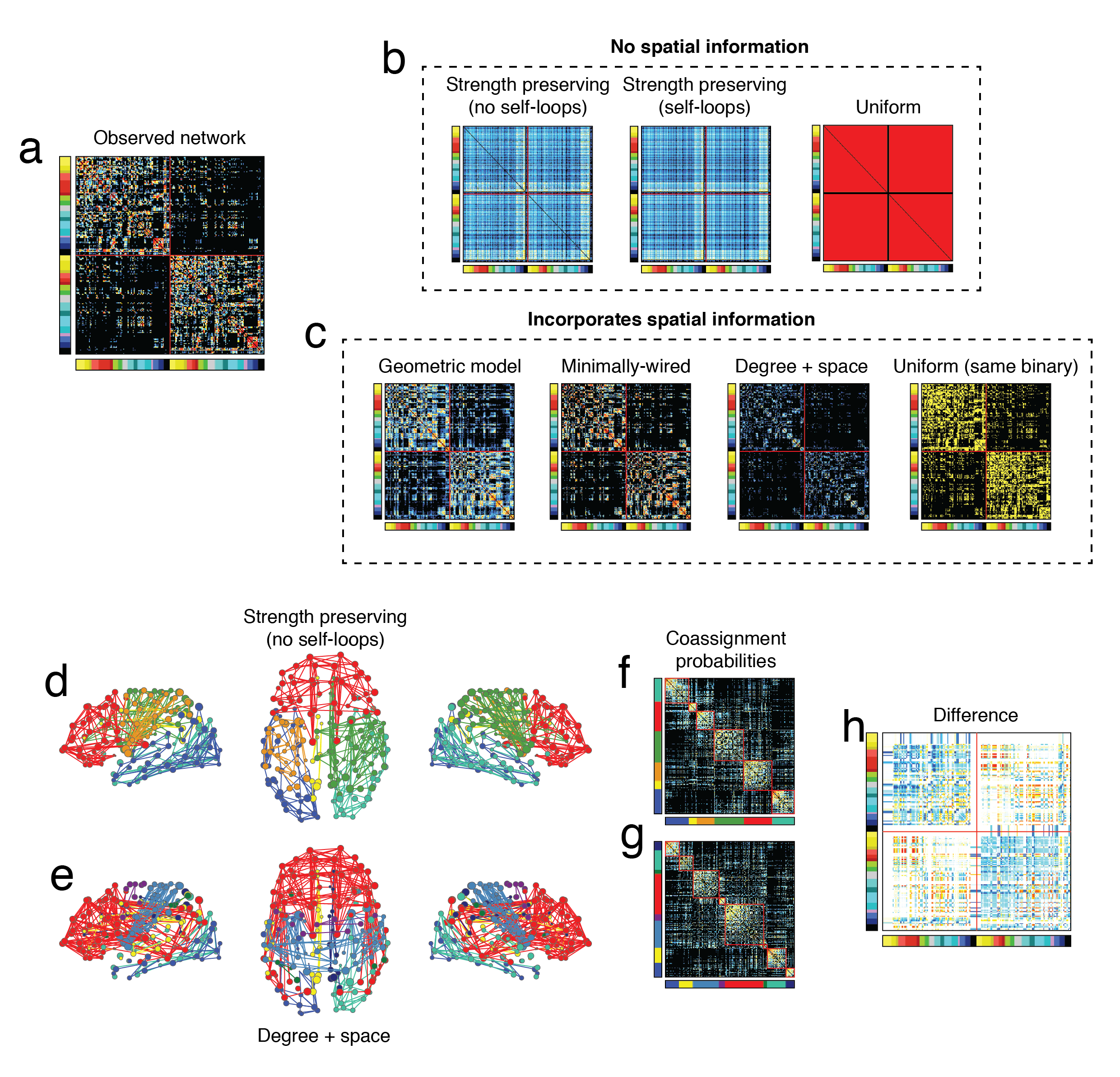}
	\caption{\textbf{Alternative null models for modularity maximization.} Modularity maximization requires that a user specify both the observed and expected weights of connections. (\emph{a}) An example observed network (in this case an empirical, group-averaged structural connectivity network). (\emph{b}) Common null models do not preserve spatial information. (\emph{c}) There exist many null models that preserve spatial information (and in some cases other information about the network, e.g. nodes' degrees). Note that, in general, models in which spatial information is preserved tend to be more similar to the observed network. This is evident even upon visual inspection. (\emph{d}) Consensus community labels obtained using a null model in which spatial information is absent. (\emph{e}) Consensus community labels obtained using a null model that preserves spatial information along with nodes' degrees. Panels \emph{f} and \emph{g} show module co-assignment matrices for both, revealing that they both generate ``internally dense and externally sparse'' assortative communities. (\emph{h}) The element-wise difference between co-assignment probability matrices, however, reveals that the strength-preserving model tends to group nodes in the same hemisphere into the same community, while the spatial model allows nodes in different hemispheres to belong to the same community.} \label{figure+1}
\end{figure*}

Most commonly, modularity maximization is used to discover the modular structure of a single network. In neuroscience, this network might represent the structural or functional connectivity from a single subject or it may be a composite (average) of connectivity data from many individuals \cite{betzel2019distance}. This is arguably the simplest modularity maximization problem. The aim is to discover modules by maximizing $Q$, which is a function of two variables: $A$, the connectivity matrix from the network of interest, and $P$, the expected weight of connections. All the user needs to do is define $P$ and run the optimization heuristic.

\subsection*{Null models for brain networks}

What is the appropriate choice for $P$? Consider, for a moment, two brain regions labeled $i$ and $j$. Maybe your network of interest is structural, and the voxels along the white-matter tract between $i$ and $j$ have some average fractional anisotropy, which represents the observed weight of the connection. What is the expected FA of that connection? What is the null (connectivity-based) model that gave rise to that expected weight? And further, what network characteristics does the null model consider? Does it take into account the spatial embedding of the connection, and does it avoid rendering self-loops? Perhaps the network of interest is derived from functional imaging data, and the edges represent co-activity patterns between regions. What null model would account for the statistical relationships of this network?

\subsubsection*{Configuration models}

These are questions that the user must confront when performing modularity maximization. However, in most applications the entire enterprise of selecting a null model and defining $P$ is sidestepped. Rather than thinking about what might be the appropriate model, the user lets the software implementation decide for them (if the software even gives the user the option of deciding), usually defining $P$ based on a ``configuration model''. The most common model of this class is one that generates null networks whose nodes have the same degree and strength as in the original network, but where connections are otherwise formed at random (Fig.~\ref{figure+1}b). Under this model, the expected weight of a connection is given by:

\begin{equation}
P_{ij} = \sum_{ij} \frac{k_i k_j}{2m}
\end{equation}

\noindent where $k_i = \sum_j A_{ij}$ is the degree of node $i$ and $2m = \sum_j k_j$ is the total number of edges in the network (or node strength and total weight if the network's edges are weighted).

Intuitively, by selecting this model the user is comparing their observed network against the set of all possible networks with precisely the same degree/strength sequence. This is an important point; the configuration model preserves a set of features of the observed network (same number/weight of edges, same degree sequence), while randomizing others. By preserving these features, specifically nodes' degrees, the configuration model also guards against the possibility of two nodes being connected to one another and assigned to the same module simply because they form more connections.

While the configuration model is a good general null, is it an appropriate null generator for brain networks? If not, what network features should be preserved in neuroscience-specific null model? How do we efficiently generate randomized networks with those features?

\subsubsection*{Incorporating spatial relationships}

One of the key organizational features of brain networks is the dependence of connections and their weights on nodes' spatial embeddings. In general, the further apart two brain regions are from one another, the less likely they are to be connected and the weaker the weight of the connection. This is especially true in the case of SC. In other words, connections are not formed independently, but according to the geometry in which the brain is embedded. These dependencies are critical for viability; longer and thicker connections are more costly, requiring more material to form and more energy to maintain and use for signaling. Moreover, they must be compact enough to fit within the skull. Therefore, over time, brains have come to adopt wiring diagrams that favor the formation of short (and less costly) connections. A brain that violates this soft constraint and forms many long connections quickly becomes too materially and metabolically expensive, exceeding the limits of viability.

This is the chief problem with configuration models. In practice, these models assume that connections are formed randomly and independently; the only constraint is that the random networks' nodes have the same degree/strength as the original network. This process, however, incorporates no information about spatial dependencies; networks generated using the configuration model contain many times more long distance connections and their total cost (measured as the summed length of their connections) far exceeds what is observed in any empirical brain network \cite{gollo2018fragility}. The configuration model, because it fails to preserve this essential feature of brain networks, can be viewed as far too liberal and almost a straw man when it comes to comparisons with empirical data.

Fortunately, generating randomized networks that preserve a network's wiring cost is a simple process. In fact some of these algorithms can simultaneously preserve cost and node degree (although it is much more difficult to preserve node strength as well (Figure \ref{figure+1}c). 

One strategy for preserving spatial relationships is to preserve the binary structure of the observed network but to randomly assign weights to the edges \cite{roberts2016contribution}. This model naturally preserves each node's degree as well as the total cost of wiring. In addition, the model also sets each edge's expected weight to an identical value equal to the mean weight of all connections in the original matrix.

A second possibility is to construct a ``minimally-wired'' model. In this model, one calculates the mean interregional distance between every pair of nodes and places connections between the $K$ shortest distances, where $K$ is the number of edges in the original network. Weights can then be introduced through rank-preserving procedures. Note that this model does not preserve nodes' degrees.

A third possibility is to parametrically model the effect of distance. In these models, we define a function that monotonically varies as a function of distance (or possibly other variables), e.g. $P_{ij} = \exp(-\gamma D_{ij})$ or $P_{ij} = D_{ij}^{-\eta}$. Here, $P_{ij}$ is the probability that nodes $i$ and $j$ are connected and $D_{ij}$ is the distance between those nodes. As long as $\gamma > 0$ and $\eta > 0$, the probability of forming long connections will be smaller than short connections. Using these models, connections can be populated using two methods -- first, one can calculate the probability for every pair of nodes and then, for every $\{  i,j  \}$, flip a biased coin to determine whether or not a connection should be added. The variables $\gamma$ and $\eta$ can be selected so that the expected number of connections generated by the model matches that of the observed network $K = \langle K \rangle$. A second possibility is to change the equalities to relative probabilities. i.e. $P_{ij} = exp(-\gamma) \rightarrow P_{ij} \propto \exp(-\gamma)$, and iteratively add edges until $K$ edges in total have been placed. This procedure ensures that networks generated by this model have exactly $K$ connections.

So far, each of these models preserve select features of a spatially constrained brain network but never more than one. In general, preserving multiple features is a more challenging task and, for certain sets of features, the space of possible random networks generated by the null model may be small or difficult to access. However, there are several strategies for doing so. One possibility is to use a ``space-aware'' configuration model. The generic configuration model can be approximated using a rewiring or ``edge-swapping'' procedure, in which two edges between four distinct nodes are swapped, so that if the initial set of edges were formed between nodes $\{ i,j \}$ and $\{ u,v \}$, the new set could be formed between $\{ i,u \}$ and $\{ j,v \}$ or $\{ i,v \}$ and $\{ j,u \}$. This procedure can be made ``space-aware'' by forcing the new edges to have similar lengths as the two original. The result is a network whose wiring cost approximates that of of the original network and whose degree sequence is identical. This procedure also approximately preserves the relationship between edges' weights and lengths.

The methods described above work well for sparse networks where edges are generated independently. However, for networks whose edges are interrelated to one another, e.g. correlation networks and functional connectivity, this method may be inappropriate \cite{zalesky2012use}. Introducing spatial relationships into null models of functional connectivity is, therefore, more challenging and, in some ways, not as necessary. While the effect of space on SC is well-documented \cite{honey2009predicting}, functional connections reflect statistical relationships between recorded activity and not physical tracts. Consequently, there is no explicit cost associated with forming a functional connection, whereas there is a cost associated with a structural connection. Indeed, while the correlation of functional connectivity weights and distance is strong for very short connections, beyond $\approx$ 40 mm, functional connections can take on a wide range of values, including some that are exceptionally strong, e.g. homotopic partners.

Nonetheless, it may be useful to incorporate spatial information into the null model for FC. One possibility for doing so is based on quadratic norms, as in \cite{bellec2006identification}. Briefly, this procedure entails fitting a parametric model of distance-dependence to data. The model, importantly, preserves transitive properties of correlations, as does other similar models that fit spatial properties of correlation matrices \cite{burt2020generative}.

Another strategy is to incorporating spatial relationships into correlation matrices at the level of time series. For instance, one recent paper proposed a method for generating a distance-dependent correlation matrices through a reweighting of uncorrelated regional time series \cite{esfahlani2020space}. For every region, its time series was defined as a weighted sum of all other regions' time series, where the weights were inversely proportionally to the Euclidean distance between regional centroids. Accordingly, nearby regions contributed more than distant regions. As a result, the elements of the resulting correlation matrix decayed monotonically as a function of distance and, because the weights from the summation were parameterized, the rate of decay could be modulated to match the sharp decay in FC over the first 40 mm seen in empirical FC data.

Spatially-informed null models incorporate an element of realism not observed in the configuration model. At the very least, they show that the choice of null model will, in general, impact the character of the detected communities and their subsequent interpretation. This was demonstrated in two recent papers \cite{betzel2017modular, esfahlani2020space} in which configuration models were explicitly compared with models that encode spatial information. In \cite{betzel2017modular}, the authors fit a spatial generative model to SC data and used the model to define the expected weights of connections. The resulting modularity matrix emphasized connections whose existence would not be anticipated had the network been generated according to cost-reduction principles alone. The communities detected using the model reflected this, exhibiting broader spatial extent than those detected with the configuration model, including instances of spatially disjoint communities, a feature that is usually never observed using the standard configuration model.

More recently, \cite{esfahlani2020space} applied a similar model to FC data from the Human Connectome Project \cite{van2013wu}. The authors demonstrated that, after correcting for spatial relationships, nodes in sensorimotor systems exhibited increased participation coefficient, suggesting that spatial relationships may obscure the centrality of some regions. Then, using the spatial model as a null model for modularity maximization, the authors detect communities and show that they, in general, disagree with previously delineated communities \cite{schaefer2018local}.

\subsubsection*{Null models for signed matrices}

Another way that the modularity equation can be (and often is) reimagined, is to make it compatible with signed matrices. This is a necessity when dealing with functional connectivity data where weights are estimated using Pearson correlation. The result is a matrix in which every node is connected to every node by a connection of some magnitude, oftentimes of mixed valence (positive and negative connections).

Naively, one might ask why the standard configuration model is inappropriate. Recall that the configuration model calculates expected connection weights based on the (weighted) degrees of nodes. In the case of correlation matrices, every node has identical degree and, because at least some fraction of connections are negative, a simple summation of their weights can return a misleading value for a node's weighted degree (because positive and negative connections can offset one another).

Some of the earliest strategies for dealing with signed matrices is to construct a modularity function that deals separately with the positive and negative elements. That is, given a connectivity matrix, $A$, to create two distinct matrices, $A^+$ and $A^-$, comprised of the positive and negative connections only. Then, to use the configuration model to estimate the expected positive and negative weights of connections as $P_{ij}^{\pm} = \frac{k_i^{\pm}k_j^{\pm}}{2m^{\pm}}$ and, from these elements, two modularity matrices: $B^{\pm} = A^{\pm} - P^{\pm}$. Intuitively, communities should correspond to groups of nodes that are cohesive (mutually positively correlated). To realize this intuition, we can define the following modularity \cite{gomez2009analysis}:

\begin{equation}
Q^{signed} = \sum_{ij} [B_{ij}^+ - B_{ij}^-] \delta(\sigma_i \sigma_j).
\end{equation}

\noindent Optimizing this modularity returns groups of nodes that are more correlated with one another than expected by chance and less internally anticorrelated than expected by chance.

Note that in this example, the contributions to $Q^{signed}$ by positive and negative connections are weighted equally. For networks containing approximately the same number of positive/negative connections this may be acceptable. However, for networks with imbalances between positive/negative connection weights, this is more of a problem. Accordingly, recent variations on this general model have aimed to balance these contributions by weighting the positive and negative terms differently \cite{rubinov2011weight}.

A second, possibly more important limitation of this approach, is that the configuration model assumes independence between edges. For structural connectivity where there is no mathematical constraint on connection weights, this is not an issue (although as noted previously, the brain's intrinsic geometry imposes a different set of soft constraints). However, for correlation networks like functional connectivity, connection weights are dependent on one another. That is, given the connection between nodes $\{i,j\}$ and $\{j,k\}$, we can place bounds on the weight of the connection between nodes $\{i,k\}$. Accordingly, our null model should also satisfy this property.

There exist several strategies for doing so. One possibility is construct a generative model that operates at the level of time series and create synthetic data, whose correlation matrix can be computed and will satisfy the transitive relationships observed in any correlation matrix. This model can then be sampled many times, the matrices averaged, and an expected value for each connection estimated. This, however, is computationally expensive, and requires the user to construct the appropriate null model, which may not always be easy to do in the time domain. Another possibility is to use approaches like the Hirschberger-Qi-Steuer (HQS) model \cite{hirschberger2007randomly,zalesky2012use} to generate admissible covariance matrices whose elements follow a specific distribution.

More recently, another study pointed out that the configuration model has a complicated interpretation in the context of correlation networks (appropriateness aside) \cite{bazzi2016community, macmahon2013community}. As an alternative, the authors suggested that a uniform null model may be appropriate for correlation matrices. The uniform null model is one in which it is assumed that every element in the network is mutually correlated with the same magnitude. That is, $P = \mathbf{1} \cdot \gamma$, where $\mathbf{1} \in \mathbb{R}^{N \times N}$ is a matrix whose elements are all 1 and $\gamma$ is the magnitude of mutual correlation. This model performed well in benchmark analyses and has the added advantage of not suffering from the resolution limit \cite{traag2011narrow}.

In this section, we highlighted several strategies for modifying the modularity equation to make it compatible with correlation matrices. Like strategies for incorporating spatial information, these strategies yield more conservative, but ultimately more realistic, null models that respect the properties of correlation matrices and functional connectivity. Failing to preserve low-level features of a correlation matrix amounts to a straw man argument and increases the risk of detecting spurious or misleading community structure.

\subsection*{Condition and group differences}

In the previous section, we highlighted potential null models for modularity maximization and noted that each can be viewed as a distinct null hypothesis. From this perspective, modularity maximization can be viewed analogously to the null- or point-hypothesis testing; comparing an observation with some intuition of chance and measuring a summary statistic as output (it is, of course, not truly the same). Whereas the output in traditional null hypothesis testing is a p-value, In the case of modularity maximization, the output is a mapping of nodes to communities and a $Q$ score. Here, the similarity between modularity maximization and hypothesis testing is by analogy only. However, is there a way to use modularity maximization to perform some version of null hypothesis testing, for instance in case-control studies to compare a group of patients with controls or to contrast FC acquired during one task with FC acquired during another? In this section, we explore this question.

\subsubsection*{Comparing FC across tasks}

\begin{figure*}[t]
	\centering
	\includegraphics[width=1\textwidth]{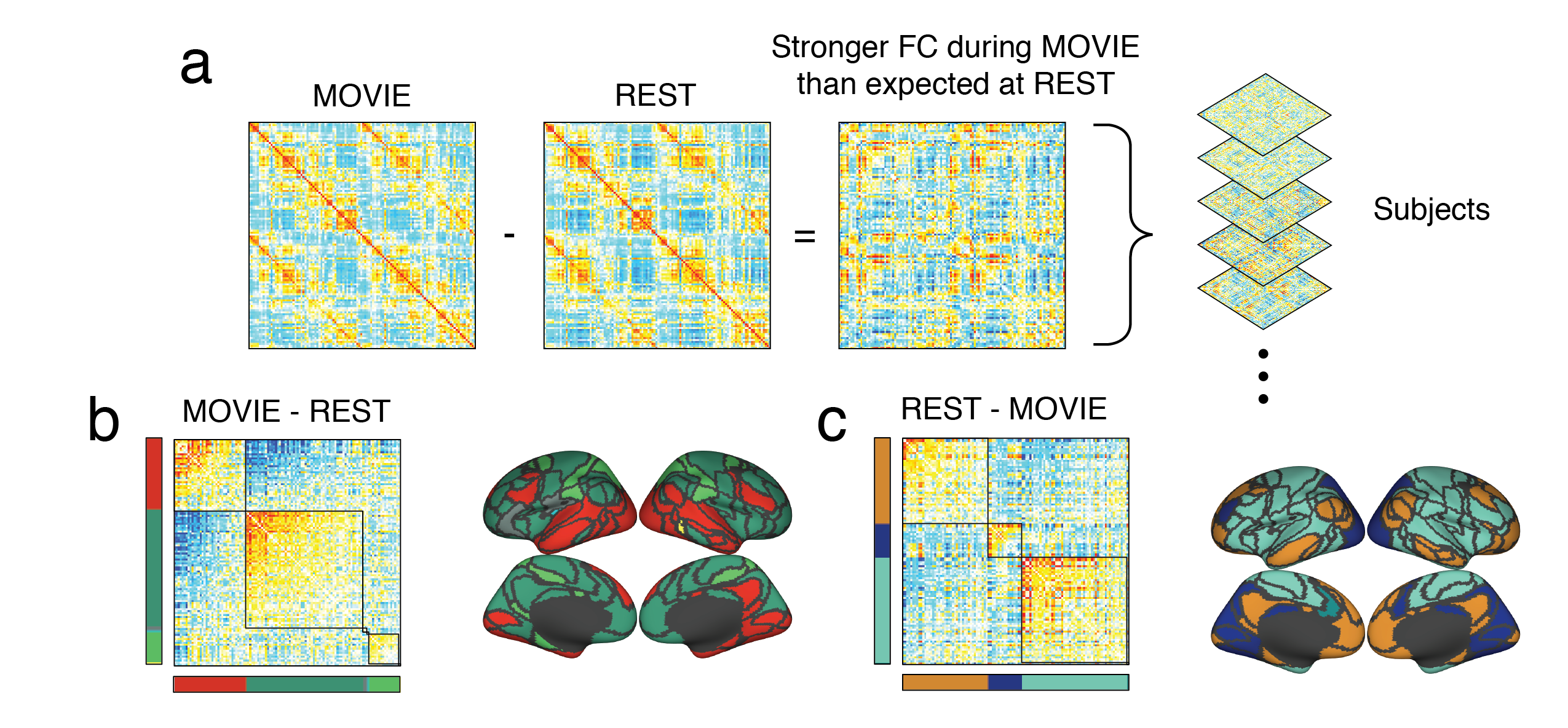}
	\caption{\textbf{Schematic illustrating how modularity maximization could be used to detect group/condition differences.} Modularity maximization works by comparing the connectivity that is observed with what is expected, but is flexible to how those matrices are defined. (\emph{a}) Here, we construct a modularity matrix by treating the observed network to be FC estimated during movie-watching while the expected network is FC estimated at rest.(\emph{b}) This modularity matrix can be optimized directly to discover communities of nodes that are more strongly correlated with one another during movie-watching than at rest. (\emph{c}) By flipping the sign of the matrix, we can discover groups of nodes that are more strongly correlated at rest than during movie-watching. Note that, in principal, this procedure must be accompanied by a statistical model to reduce the possibility of false-positives, as modularity maximization will always find communities in data, even if none exist.} \label{figure+2}
\end{figure*}

One way that modularity maximization could be extended to test distinct null hypotheses is in the comparison of FC acquired during different task states with FC at rest. For example, how does the brain's community structure change when performing a working memory task compared to, say, rest? The conventional response to this question is that the changes are subtle but systematic \cite{cole2014intrinsic}. Resting state modular organization is an optimized state, where metabolic demands are minimized, allowing the brain to rapidly reconfigure in service of any forthcoming task-related processing goal \cite{wig2017segregated}. Thus, FC reconfigures only slightly from rest-to-task and their connection weights are highly correlated. What if we reframed the question slightly, such that we consider ``rest'' to be our null or baseline condition -- the $P$ in the modularity expression -- and working memory FC to be our observed network -- the $A$ matrix (Fig.~\ref{figure+2})? In this case, we could construct a modularity matrix:

\begin{equation}
B = FC_{task} - FC_{rest}.
\end{equation}

\noindent We could then use this matrix to recover groups of nodes whose mutual connections strengthen during the working memory task compared to rest. Interestingly, the sign of this matrix could also be flipped, so that $B' = FC_{rest} - FC_{task}$. In this case, the detected modules now reflect groups of nodes whose connections are stronger at rest than during the working memory task.

\subsubsection*{Comparing connectivity between groups}

If we are willing to consider the hypothetical scenario described above, then there are other interesting ways that modularity maximization can be modified to accommodate other hypotheses and research questions. For instance, we could use it to discover system-level differences in case-control studies, e.g. where some clinical or treatment population gets compared against a control group \cite{alexander2010disrupted}. In this case, we can regard the connectivity of the typical control subject as our expected connectivity pattern; the configuration of connection weights that represent a neuro-typical subject. The clinical group, on the other hand, represents what we actually observe. Using these two components, we can construct a modularity matrix:

\begin{equation}
B = FC_{clinical} - FC_{control}.
\end{equation}

\noindent Optimizing modularity based on this matrix will result in communities whose mutual connections are stronger in the clinical condition than they are in a neuro-typical individual. We note, of course, that we could test the opposite hypotheses (what are the clusters of nodes that are stronger in controls than in the clinical group?) by simply flipping the sign of the modularity matrix or equivalently by subtracting the clinical FC from that of the controls.

This extension of modularity maximization to test for group-level differences in meso-scale structure has yet to be explored. We note, however, that this approach can be viewed as analogous to other methods, most notably the \emph{network-based statistic} (NBS), which uses an element-wise statistical map (e.g. correlations or group differences of edge weights), applies a statistical threshold, and extracts the largest supra-threshold component \cite{zalesky2010network}. With modularity maximization, we instead uncover subsets of nodes with changed connectivity patterns. Importantly, modularity maximization makes no \emph{a priori} assumption about the size of these subsets. NBS, on the other hand, presumes that the group-level effect gets manifested in the largest connected component. 

Using modularity maximization to detect clusters of nodes whose connectivity changes with condition requires some additional caveats. For instance, it requires developing methods for controlling false positives. In other words, modularity maximization will \emph{always} return a partition of a network into communities. If there exists element-wise differences in $FC_{clinical}$ and $FC_{control}$ or $FC_{rest}$ and $FC_{task}$, even in the absence of true clusters, then modularity maximization could spuriously detect modules, i.e. find signal in noise. Controlling for this possibility requires exquisitely accurate estimates of the connectivity matrices (likely not possible) or a carefully constructed null model. In the case of null models, one possibility would be to estimate the group-level matrices using random permutation of group assignments. That is, one could randomly assign some of the clinical data as controls and \emph{vice versa}, optimize the resulting modularity, and compare the detected communities and their quality with those generated using the true condition labels. This very issue, in fact, arises in applications of NBS, and is addressed using the same permutation strategy. As we will also illustrate in the next sections, another solution to mitigate false positives and facilitate the tracking of changes across conditions is adopting a multi-layer framework.

\section*{Multi-layer modularity maximization}

The previous two sections focused on strategies for modifying the modularity quality function by directly changing the expected weights of connections -- the $P$ matrix in the equation $B = A - P$. These changes transform modularity maximization from a simple community detection algorithm, into a framework for investigating some of the central questions in neuroscience, including the comparison of groups and experimental conditions.

These analyses, however, also expose a feature (or perhaps limitation) of modularity maximization that is shared by most other community detection algorithms. Namely, these algorithms are designed to estimate communities for one network at a time. If a network's structure varies across time (e.g. time-varying FC \cite{lurie2020questions}), if we want to consider multiple subjects in a cohort \cite{betzel2019community} or compare communities across different connection modalities (e.g. SC, FC, structural covariance, gene co-expression, etc. \cite{bentley2016multilayer}), the standard modularity function is poorly suited for doing so. To detect modules in datasets made up of multiple connectivity matrices requires that we extend modularity maximization further.

%This is, of course, another fundamental problem in neuroscience. Consider a cohort of tens, hundreds, or even thousands of individuals, each of which is associated not only with their own connectivity matrix, but also a series of attributes and traits. We may be intesre

When we consider network datasets comprised of multiple connectivity matrices, there are two general strategies by which they can be analyzed. On one hand, each of the networks can be  analyzed independently from all of the others. In the case of a multi-subject cohort, this involves estimating modules for each of the subjects and then comparing them at the end. However, this strategy quickly leads to problems aligning communities from one subject to another. That is, when can we say that communities detected using data from subjects A and B are realizations of the same community? With the exception of trivial cases (where the communities contain identical sets of nodes), this problem becomes challenging and is not easily nor unambiguously resolved, often requiring that the user introduce an additional heuristic.

The second strategy is to simply average the multiple connectivity matrices, forming a single composite matrix. While this strategy allows us to use well-developed community detection methods (or even the variants described in the previous two sections), it also results in a loss of information; by averaging the weights of connections across many matrices, we emphasize their average features, but lose all information about their variability.

Clearly, both strategies present challenges and tradeoffs; the first is capable of capturing inter-subject variation, but has challenges aligning communities from one individual to another. The second approach is more parsimonious and circumvents the alignment problem, at the expense of information about how connectivity patterns (and communities) differ or vary from one matrix to the next. Which should we choose then?

\begin{figure*}[t]
	\centering
	\includegraphics[width=0.6\textwidth]{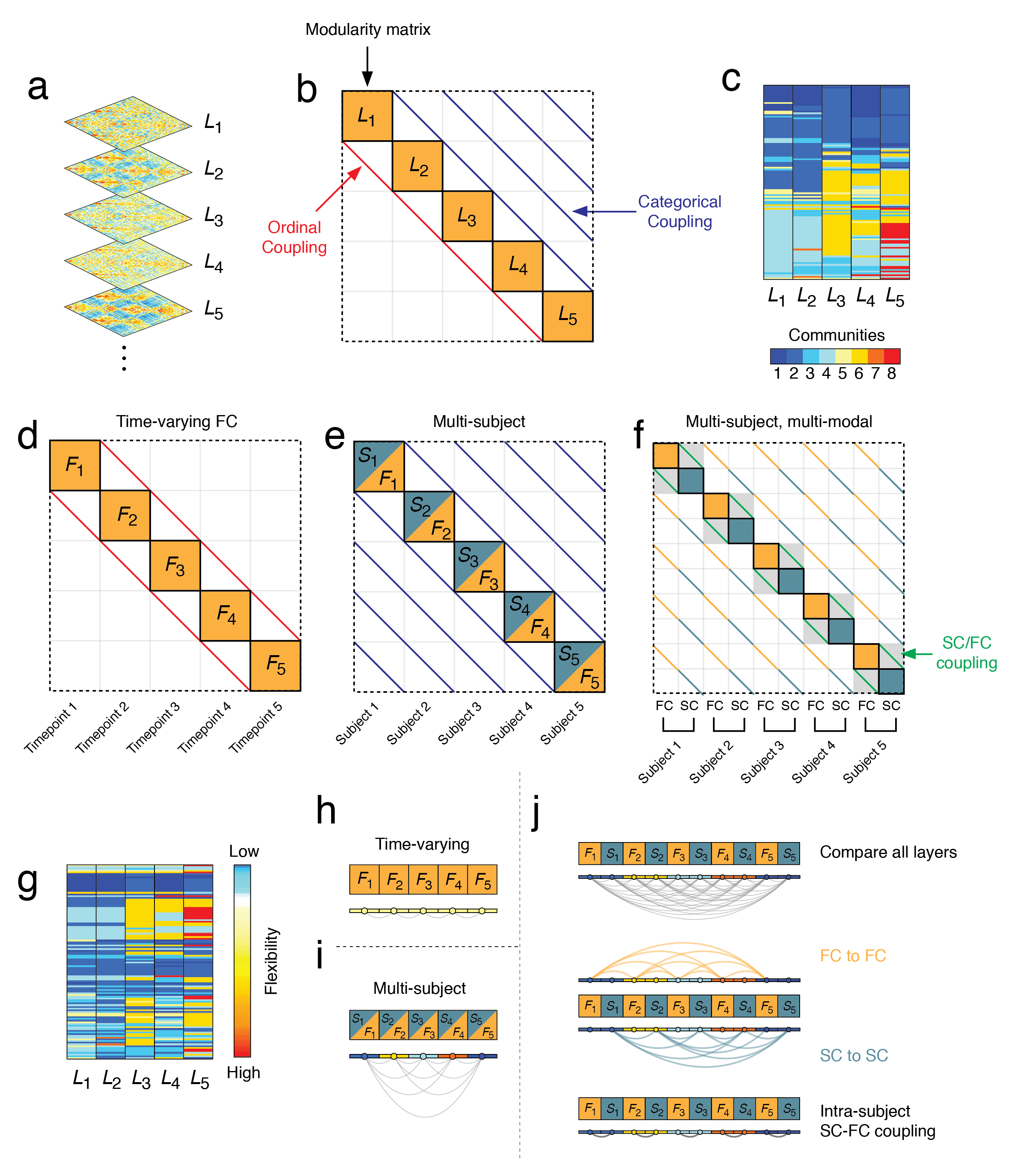}
	\caption{\textbf{Multi-layer modularity framework.} (\emph{a}) Multi-layer modularity operates on network datasets comprising many distinct connectivity matrices or ``layers''. (\emph{b}) To detect communities, each connectivity matrix (or more accurately, its modularity matrix) is embedded along the block diagonal of a modularity tensor (shown flattened here). Then, a small amount of weight is added between each node and itself across layers. When networks have some temporal ordering, these additional weights may only connect adjacent layers. For networks without temporal order, weights can be added all-to-all. (\emph{c}) The modularity tensor is then passed to an optimization algorithm to simultaneously partition nodes in each layer into communities. The advantage of this approach is that community labels are preserved across layers, making it possible to directly compare nodes' community assignments across layers. In previous applications, multi-layer modularity has been applied to time-varying networks (\emph{d}) or multi-subject datasets (\emph{e}). In these diagrams, orange and green denote modularity matrices from FC and SC, respectively. However, the framework allows for more complicated configurations, including incorporating SC and FC simultaneously, which allows for the inclusion of a new class of inter-layer coupling (SC to FC; panel \emph{f}). (\emph{g}) Multi-layer partitions also make it possible to directly compare communities across layers using a measure of flexibility -- how frequently assignments change from layer to layer. This diagram depicts the same communities as in panel \emph{c} but with rows ordered in terms of flexibility. With different multi-layer tensors, it is possible to calculate different versions of flexibility. In time-varying networks flexibility is calculated by comparing communities only between temporally adjacent layers(\emph{h}) while in multi-subject datasets, it makes sense to compare communities between all pairs of layers. (\emph{j}) In multi-subject/multi-modal datasets, we can calculate different categories of flexibility. For instance, in addition to an all-to-all comparison, one could calculate the flexibility between FC layers or SC layers. One could also calculate SC-FC flexibility by comparing SC and FC layers within subjects.} \label{figure+3}
\end{figure*}

A third strategy for approaching this problem is to take advantage of multi-layer networks \cite{kivela2014multilayer} (Fig.~\ref{figure+3}a). Whereas an individual connectivity matrix is a two-dimensional object that encodes pairwise connections among a fixed set of elements, multi-layer networks can be conceptualized as a ``stack'' of these network ``slices'' or ``layers'', forming a three-dimensional multi-layer object. Multi-layer networks are used throughout network science \cite{de2017multilayer, vaiana2018multilayer} and recently have begun to be used widely within network neuroscience, where they are typically used to study dynamic FC \cite{bassett2011dynamic}, where network organization is modeled through a series of time-varying connectivity matrices.

One of the principal advantages of multi-layer networks is that they can aggregate many connectivity matrices into the same (multi-layer) network model, enabling those networks to be analyzed simultaneously without a loss of information. Among the many possible tools for analyzing multi-layer networks is a multi-layer analog of modularity maximization \cite{mucha2010community, bazzi2016community}. Although we refer to it as an analog of the traditional single-layer modularity maximization, it is in fact \emph{identical} to the single-layer method, operating on the exact same principles and formulation.

Multi-layer modularity maximization works by creating a special kind of modularity matrix. Along the diagonal of this matrix are modularity matrices from individual layers; one for each of the $K$ single-layer networks we want to analyze (Figure \ref{figure+3}b). If we were to submit this matrix to any modularity maximization algorithm nodes from different layers will \emph{always} be assigned to different modules. That is, node $i$ in layer $s$ would never appear in the same community as node $i$ in layer $t \ne s$. This is because there exist no cross-layer groupings of nodes that would lead to an improvement in $Q$. In other words, there is no practical difference between this approach and the case where we analyze each network independently.

However, we can change this fact by, once again, directly modifying the modularity matrix. Specifically, we can add a small amount of weight, $\omega$, linking node $i$ in layer $s$ to itself in other layers\footnote{ note that the inter-layer coupling does not have to be in this form -- it could, in principle, be added as any pattern of off-diagonal coupling between modularity matrices}. In fact, we can do this for \emph{every} $i \in [1,\ldots, N]$ \cite{puxeddu2019optimal}. The effect of the $\omega$ parameter is that there now exists scenarios where merging nodes from different layers can lead to improvements in modularity, i.e. $\Delta Q > 0$. In a way similar to $\gamma$, we can tune $\omega$ to detect communities more or less consistent across the slices of the multi-layer networks. Suppose we have a multi-subject network. It can be that all the subjects belong to a specific category, or are performing the same task, and we want to identify a common community organization that can be associated with a category or task. In this case, we could use high $\omega$ values and encourage the algorithm to find common structures across subjects. On the other hand, if we want to investigate inter-individual variability, or we have a more heterogeneous group of subjects to deal with, we are more likely to chose a low $\omega$, that will highlight features unique to the subjects. Multi-layer modularity optimization with $\omega$-tuning has been shown to be more robust to noise with respect to the single-layer case applied to each slice, and this is true in both conditions where we investigate steady structure within the ensemble of networks,  or non-steady ones \cite{puxeddu2021comprehensive}

Because community labels are preserved from one layer to the next, we obviate the need to align communities. Comparing communities in different layers is as simple as assessing whether their labels are or are not identical. When averaged across all comparisons, this measure is referred to as \emph{flexibility} \cite{bassett2011dynamic, bassett2013robust, braun2015dynamic, braun2016dynamic, gerraty2018dynamic, chai2017evolution} (Figure \ref{figure+3}g). This approach requires no averaging procedure and so no information about individual connectivity matrices is lost in the process.

Although applications of multi-layer community detection are becoming increasingly common within the network neuroscience community, there remain (simple) opportunities for innovation. The following section will continue our discussion of different ways that multi-layer modularity has been used in neuroscience while highlighting some of the possibilities for improvement.

%We can think about this problem in terms of multi-layer networks. A single connectivity matrix is a two-dimensional object that encodes pairwise connectivity among a fixed set of elements. We can conceptualize situations where there are multiple connectivity matrices as a three-dimensional ``stack'' of these two-dimensional ``slices'' or ``layers''. To study the community structure of these matrices jointly, we need to perform \emph{multi-layer community detection}. Like single-layer community detection, there are many methods for estimating communities in multi-layer networks. Yet again, one of the most population is the multi-layer extension of modularity maximization.

%In many ways, multi-layer modularity maximization is not an extension of the single-layer method. Rather, it is \emph{identical} to the single-layer method and, as in the previous two sections, and is made compatible with multi-layer networks by directly modifying the modularity matrix. \textcolor{red}{YOU LEFT OFF HERE}

%\textcolor{red}{would be good to add a table that summarizes all of things we're proposing doing with modularity maximization}

%A final extension of modularity maximization involves clustering multiple networks simultaneously. This is accomplished by representing each network as a layer in a multi-layer network object. This multi-layer network and all of its layers are clustered in a single processing step. Whereas clustering layers independently is also possible -- and sometimes more efficient, computationally -- it results in the challenging problem of creating a mapping of communities from one network to another. This problem is, in fact, ill-posed, and not uniquely solvable. The multi-layer model, on the other hand, uses a labeling system of communities that is consistent across layers, making it trivially easy to map communities from one subject to another.'

\subsection*{Applications to time-varying connectivity}

Presently, applications of multi-layer modularity maximization to neuroimaging data have been narrow and largely restricted to dynamic or time-varying networks. In these studies, each layer represents a functional network that describes connectivity at a given point in time, e.g. a narrow window in time during a scan session \cite{bassett2011dynamic,telesford2016detection,yang2021measurement} or a particular developmental stage \cite{yin2020emergence,puxeddu2020modular}.

In these examples, multi-layer modularity maximization makes it possible to track the formation, evolution, and dissolution of community structure across time. This is accomplished by adding weak connections between a node and itself in successive layers and using algorithms like the Louvain algorithm to optimize this multi-layer version of $Q$. The outcome, as noted earlier, is a series of community labels that are propagated across layers. Because these labels are preserved from one layer to another, comparing communities across layers is trivial. 

Again, we want to emphasize that this feature -- the correspondence of community labels across layers -- is one of the principal advantages of multi-layer modularity maximization. Matching communities from one partition to another is, in general, non-trivial for single-layer networks and in some cases is ill-posed. Consider, as an example, two networks, made up of four nodes. In network 1, we find those nodes all assigned to the same community $\sigma_x = \{ A , B , C , D \}$. In network 2, we find them split evenly into two communities $\sigma_y = \{  A , B \}$ and $\sigma_z = \{  C , D \}$. How is $\sigma_x$ related to $\sigma_y$ and $\sigma_z$? Does $\sigma_x$ fracture into two novel communities or does it continue on through one of either $\sigma_y$ or $\sigma_z$? If so, which of the two represents the continuation and which represents a novel community? Or are they both novel?

Addressing these types of questions can be challenging when communities are detected using single-layer modularity maximization. In all but the most trivial cases, matching communities across partitions requires the inclusion of some additional heuristic, e.g. to ``break ties'' like those described above. Multi-layer modularity maximization, on the other hand, incurs a greater up-front computational cost (increased memory to store multilayer networks, greater runtime to estimate communities), but automatically tracks communities across layers, effectively solving the matching problem. In doing so, it facilitates the calculation of a large suite of metrics, including \emph{flexibility} (Figure \ref{figure+3}g), which measures how often a node changes communities from one layer to the next. Flexibility, in particular, has been used in many contexts, and has been linked to learning rate \cite{bassett2011dynamic}, mood and affect \cite{betzel2017positive}, clinical status \cite{braun2016dynamic}, and executive function \cite{braun2015dynamic}, among others.

\subsection*{Applications to multi-subject, multi-task, and dense sampling datasets}

While most applications of multi-layer modularity maximization and flexibility analysis have been to time-varying FC, they have also be applied in other contexts. For instance, to investigate differences in modular structure across task state and across subjects \cite{betzel2019community}. Just like with time-varying connectivity, multi-layer modularity can be used to detect modules in networks where layers represent connectivity estimated under different conditions (e.g. different tasks in the scanner) or even different individuals (Figure \ref{figure+3}e). As with applications to time-varying connectivity, flexibility analyses can be used to characterize changes in community structure across layers. In this case, however, the interpretation changes -- flexibility becomes a measure task-induced reconfiguration or inter-subject stability of communities. Note that one could also combine these two approaches and use multi-layer modularity to investigate changes in modular structure in multiple brains across time. This approach might be especially useful in studies using naturalistic stimuli \cite{sonkusare2019naturalistic, betzel2020temporal, finn2021movie}, or hyperscanning experiments, in which brain activity is recorded simultaneously from two (or more) individuals while interacting with one another \cite{babiloni2014social, dumas2011social}.

Interestingly, flexibility can be broken down into components by comparing specific pairs of layers. Suppose, for instance, that a dataset was comprised of a control and clinical population. You could measure the flexibility (community dissimilarity) separately between all pairs of control and clinical subjects and then within both populations. These values could then be evaluated further using basic statistics to compare the variability of communities within and between populations. This approach could also be especially powerful in dense-sampling studies, where subjects are scanned many times in order to better characterize their subject-specific organization. In these types of studies, one could treat connectivity derived from repeated scans of many subjects and conditions and subsequently calculate different versions of flexibility. For instance, one could calculate flexibility only comparing communities within subjects to estimate the within-subject variability of communities. This would allow a user to trivially generate spatial maps of regions and the baseline variability of their community assignments within subjects. The same could be done between subjects, between conditions, and for different combinations. In spirit, this flexibility analysis is similar to other studies that have examined variability of FC patterns to ascertain what factors account for their variability \cite{gratton2018functional}.

\subsection*{Multi-layer community detection for multi-modal datasets}

There are yet other ways that multilayer modularity maximization could be further extended. For instance, one interesting application involves treating different connectivity modalities as layers \cite{betzel2019structural,bentley2016multilayer}. For instance, one could assemble a multi-layer network from structural and functional connectivity, along with morphometric and transcriptomic similarity matrices and cluster them simultaneously. Importantly, consideration should be taken that the weights of the matrices are comparable. If one layer has much stronger weights than the others, it will tend to contribute disproportionately to the multi-layer modularity and impact estimates of modular structure from other layers.

Looking even further into the future, one could also imagine jointly detecting communities using multi-modal data from many different subjects (Figure \ref{figure+3}f). A clear example is the case where one has a dataset of SC and FC matrices from many different individuals. As before, ensuring that the total weight of each layer and modality are comparable to one another is important.  For instance, while SC matrices are usually sparse and made of positive weights, FC matrices are fully connected and made of positive and negative weights. This difference can be overcome by transforming the SC matrix in a structural correlation matrix, by computing the Pearson's correlation between each pair of rows. In this way, the weights of both SC and FC would fall within the same range \cite{amico2018mapping}. The user also has the opportunity to be creative in terms of how the inter-layer coupling parameter, $\omega$, is introduced. For instance, one might imagine only coupling structural matrices to structural matrices across individuals but, within subjects, introducing a new parameter that governs the subject-specific coupling of structural and functional connectivity to one another.

When the inter-subject coupling of FC-to-FC or SC-to-SC is strong and the SC-to-FC coupling weak, optimizing the multi-layer modularity will return group-representative partitions for both SC and FC. However, as we increase the coupling between SC and FC at the subject level we also force the modularity maximization algorithm to detect modular structures that are shared between the two modalities. Again, this procedure allows for the calculating various flexibilities, including the flexibility between SC and FC, both globally (considering all subjects) and at the micro-level of individual subjects. These subject-level measures of structure-function coupling could be used as phenotypical measures of an individual and their variation tracked over the course of development, across tasks, or clinical assessments.

\section*{Outlook and concluding remarks}

%This prospective article deals with brain network communities and focuses specifically on the modularity maximization heuristic. We argue that, despite its widespread application and success, modularity maximization has, so far, been applied \MGP{\textst{in }}narrowly within neuroscience\MGP{, and without really taking into account our achieved knowledge of brain networks, or the clinical questions one wants to answer}. We highlight several simple strategies for broadening and extending the modularity maximization framework by directly modifying the modularity matrix to incorporate realistic features of brain networks, including \MGP{\textst{the incorporation of }}spatial dependenc\MGP{i}es\MGP{,} and preserv\MGP{e\textst{ation of}} other features of empirical brain networks, including statistical relationships among connections. The primary aim of this work is not to present new findings or to overturn previous observations, but to draw attention to these modifications with the hope that they will lead to new neuroscientific insight\MGP{s}.

This prospective article deals with brain network communities and focuses specifically on the modularity maximization heuristic. We argue that, despite its widespread application and success, modularity maximization has, so far, been applied narrowly within neuroscience and often without taking into account domain-specific knowledge. We highlight several simple strategies for broadening and extending the modularity maximization framework by directly modifying the modularity matrix to incorporate realistic features of brain networks, including spatial dependences between connections and the preservation of statistical relationships among connections. The primary aim of this work is not to present new findings or to overturn previous observations, but to draw attention to these modifications with the hope that they encourage users to creatively explore modifications of modularity maximization, leading to new neuroscientific insight.

Community structure is omnipresent across networks \cite{newman2012communities}. Detecting this structure and designing new algorithms to do so is of great interest to practitioners and theoreticians, alike \cite{leskovec2009community,fortunato2016community}. Like other clustering algorithms, community detection reduces the dimensionality of large complex systems, identifies patterns, uncovers functionally-related groups of nodes, diagnoses node roles, and broadly generates new insight into system organization and function, prompting new scientific hypotheses.

Community detection is also important for neuroscience \cite{sporns2016modular,meunier2009hierarchical}. Shifts in data-sharing practices have generated massive, publicly-available datasets that include imaging data from thousands of individuals \cite{van2013wu, horien2021hitchhiker}. In parallel, advances in cellular-level recordings have made it possible to record from up to millions of neurons simultaneously \cite{demas2021high}. Making sense of these data requires flexible tools for dimension reduction and for extracting neurobiologically meaningful features that can be propagated to secondary analyses.

Indeed, community detection, in general, and modularity maximization, specifically, have been applied to brain data at virtually all spatiotemporal scales, offering a convenient method for partitioning nervous systems into communities. However, in almost every instance, these applications have leveraged an ``out-of-the-box'' version of modularity maximization. On one hand, the fact that neuroscientific insight can be gleaned in this way is a testament to the universality of network science and the tools used to interrogate networks. Modularity maximization was designed to be generic and not necessarily with brain networks in mind. On the other hand, this leaves lots of room for improvement. As noted earlier, the generic version of modularity maximization features a null model that is poorly-suited for both SC an FC, where it fails to preserve basic features of brain networks or outright violates statistical truisms.

The aim of this article, however, is not to wag a finger at past applications of modularity maximization. Rather, we endeavor to highlight several strategies for extending modularity maximization so that future applications can make stronger and more nuanced scientific claims and target increasingly specific research questions. Additionally, we want to highlight the components of modularity maximization -- null models, multi-layer formulations, parameterizations -- that are fundamentally left up to the user and to encourage exploration of these components. It is our view that the utility of ``out-of-the-box'' and generic methods, although they been used to make important contributions up to this point, is limited and can only carry the field so far. Future work must be directed to link data-driven and network science methods with domain-specific knowledge from neuroscience. In the following paragraphs, we cover several topics that are within the scope of this prospective article but were not discussed in depth in the main text.

%\textcolor{red}{add some references to papers frm outside of fmri-world that used modularity... there's a japanese paper, the worm paper, etc.}

%\subsection*{The role of modularity maximization}
%
%The focus of this article has been on modularity maximization, and for good reason. It operates according to an eminently simple principle, seeking communities of nodes whose internal density of connections is maximally greater than what would be expected under a chance model. Modularity maximization is a relatively old method \cite{newman2004finding} and has been applied in virtually every scientific domain in which networks have made ingress.
%
%In neuroscience, specifically, modularity has provided a bounty of new insights, uncovering structure in networks at all scales, from networks of single cells \cite{jarrell2012connectome,lee2016anatomy,betzel2019stability} to areas \cite{betzel2020organizing} to large-scale brain regions \cite{power2011functional}.
%
%Notably, modularity maximization is just one of many methods that operate on brain network data.

%\subsection*{Limitations of modularity maximization}

The focus of this article has been on modularity maximization, and for good reason. It operates according to an eminently simple principle, seeking communities of nodes whose internal density of connections is maximally greater than what would be expected under a chance model. Modularity maximization is a relatively old method \cite{newman2004finding} and has been applied in virtually every scientific domain in which networks have made ingress. Optimizing modularity yields communities that are strictly assortative -- nodes in the same community are more likely to connect to other nodes in the same community than to nodes in other communities. This type of organization, however, is one of many possible types of community configurations. Among these are disassortative -- where nodes in the same community preferentially avoid making connections to one another -- and core-periphery structure, where a densely connected core projects to a periphery and the peripheral nodes avoid making connections to other peripheral nodes. Modularity maximization is incapable of resolving these types of organization.

However, there exist many other methods that are capable of discovering non-assortative communities. Among the most popular methods is the stochastic blockmodel, which seeks to discover groups of nodes whose connectivity patterns are similar to one another. This allows for the possibility that groups of nodes are not connected to one another but nonetheless exhibit similar connectivity profiles, e.g. core-periphery or disassortative communities \cite{betzel2018diversity, faskowitz2018weighted, moyer2015blockmodels}.

Unlike modularity maximization (and Infomap), blockmodels can flexibly detect different categories of community interactions. This is an important point; modularity and Infomap are capable of detecting only assortative communities. If the aim is to obtain an estimate of the brain's true community structure, we might expect these methods to perform reasonably well, provided those ground-truth communities are assortative. However, if the brain deviates from this type of organization, then modularity and Infomap may perform poorly and misclassify nodes' communities. Indeed, these misclassifications can further impact our understanding of the brain's organization and function. Misclassified nodes may make connections outside of the assortative community to which they are assigned, giving the impression that they are integrative hubs that span modules. A blockmodel, however, might discover a more parsimonious description of communities where these putative ``hubs'' are grouped together with other hubs and form far fewer cross-module links \cite{pavlovic2014stochastic}. Additionally, blockmodels are, effectively, generative models of a network and serve as a powerful way of performing posterior predictive checks -- i.e. simulating data from the fit model -- which can be used to assess how well the model recapitulates the observed network and to identify features that are not accounted for by the current model \cite{gelman2013bayesian}.

In general, no community detection algorithm is privileged over the others, including modularity maximization. Community detection can be viewed as an inversion of some generative function for networks that depends on nodes' latent community assignments. Different algorithms and methods effectively ``guess'' at what that function is to produce estimates of communities. However, there is almost always a many-to-one mapping of a fixed set of communities to a given network and, similarly, there may be different sets of communities that, through different generative functions, explain the observed network equally well \cite{peel2017ground}. Unless one knows the true generative function, then it is impossible to determine with certainty which method and its inferred communities is optimal. One might imagine that this issue could be addressed by cross-validating with respect to some additional data, e.g. verifying that detected structural communities are enriched for functional connectivity or gene coexpression patterns. However, this strategy suffers from the same issue, in that there may exist multiple mappings between these additional data and the detected communities. In short, these observations motivate exploring multiple community detection algorithms and not weighting the results of any single algorithm heavily, but focusing instead on communities that are shared and robust, but also points of frustration, where different algorithms fail to arrive at consensus.

Since the recent explosion of interest in networks, neuroscience has repeatedly borrowed methodology from network science proper. Small-worlds, hubs, rich-clubs, and modules are concepts that make sense for brain network data, but the tools used to detect these types of structures originated in network science, often with applications to other systems in mind, e.g. social networks. While we focus here on extending one of these tools, this represents only the first steps. A key challenge for the future is to reverse this directionality and to develop analysis tools that are specific to nervous systems, taking into account their generative mechanisms, peculiarities, and domain-specific knowledge. Doing so creates bridges with other sub-disciplines in the computational neurosciences, opening up opportunities for increasingly sophisticated exploration of brain network organization.

\subsection*{Software}

Matlab code for implementing the models discussed in the paper can be found via the following link: \url{https://github.com/brain-networks/nonstandard_modularity_maximization}.

\section*{Acknowledgments}

RFB wrote initial draft of manuscript. All authors edited and revised manuscript and contributed and tested code.

\bibliography{../../bnbl_bibfile,../addbib_jif,../addbig_mgp}

\end{document}